\documentclass[11pt]{article}
\usepackage{graphicx}
\usepackage{amssymb}
\usepackage{amsmath}
\usepackage{epsfig}
\usepackage{hyperref}
\usepackage{multirow}

\begin{document}

\thispagestyle{empty}
\newcommand{\be}{\begin{equation}}
\newcommand{\ee}{\end{equation}}
\newcommand{\bea}{\begin{eqnarray}}
\newcommand{\eea}{\end{eqnarray}}
\newcommand{\half}{\frac{1}{2}}
\newcommand{\avexh}{{\langle x_H \rangle}}
\newcommand{\avexi}{{\langle x_i \rangle}}
\newcommand{\aveTb}{{\langle T_b \rangle}}
\newcommand{\aveTS}{{\langle T_S \rangle}}
\newcommand{\tabspace}{\hspace{0.4cm}}
\newcommand{\fns}{\footnotesize}
\def\lsim{\mathrel{\raise.3ex\hbox{$<$\kern-.75em\lower1ex\hbox{$\sim$}}}}
\def\gsim{\mathrel{\raise.3ex\hbox{$>$\kern-.75em\lower1ex\hbox{$\sim$}}}}

%\preprint{
% {\vbox{
% \hbox{\bf MADPH-08-1523}
% \hbox{}
% \hbox{}
% }}}

\title{Inflationary Potential from 21 cm Tomography and Planck}
\author{Vernon Barger$^{1}$, Yu Gao$^{1}$, Yi Mao$^{2,3}$ and Danny Marfatia$^{4}$\\[3ex]
\small\it $^1$Department of Physics, University of Wisconsin, 
Madison, WI 53706\\
\small\it $^2$Department of Physics, Massachusetts Institute of Technology, Cambridge, MA 02139\\
\small\it  $^3$Department of Astronomy, University of Texas, Austin, TX 78712\\
\small\it  $^4$Department of Physics and Astronomy, University of Kansas, Lawrence, KS 66045
}
%\noaffiliation
\date{}
%\vspace{0.5in}
\maketitle
\begin{abstract}

%\vspace{0.2in}

Three-dimensional neutral hydrogen mapping using the redshifted 21 cm line has recently emerged as a promising cosmological probe.  
Within the framework of slow-roll reconstruction, we analyze how well the inflationary potential can be reconstructed 
by combining data from 21~cm experiments and cosmic microwave background data from the Planck satellite.  
%We assume that a scalar field dominates the early stage of super-horizon period, and impose a prior on the number of e-folds of inflation. 
%The inflationary potential is reconstructed by kinematical slow-roll parameters. 
We consider inflationary models classified according to the amplitude of their tensor component, and show that 21 cm measurements 
can significantly improve constraints on the slow-roll parameters and determine the shape of the inflationary potential.   

%%%edited by Yi -- make sure what I massaged is correct 
%Three-dimensional neutral hydrogen mapping using the redshifted 21 cm line has recently emerged as a promising cosmological probe.  
%We analyze how the shape of inflationary potential in the early universe can be constrained by combining the data from 21 cm experiments and the CMB data from Planck.  
%We assume that a scalar field dominates the early stage of inflation, and impose a prior on the number of e-foldsings at the end of inflation.  We consider two models that correspond to high and low tensor density fluctuations, respectively.  
%The inflationary potential can be reconstructed by a class of kinematical parameters, the so-called slow roll parameters, and we show that 21 cm measurements can significantly improve the constraints on the slow roll parameters.  

%In this paper we analyze the how a 21 cm project and the CMB data from Planck satellite can determine the shape of inflationary potential in the early universe. 21 cm measurement provides significant contribution to the constraint on the slow parameters, a set of kinematical parameters that describe the inflationary potential. We assume a scalar field dominates the early stages of inflation and reconstruct inflationary potentials with up to three slow-roll parameters inside the constraints from a 21 cm+Planck Fisher matrix analysis plus a prior on the number of e-folds at the end of inflation. 

\end{abstract}

\maketitle
\newpage

\section{Introduction}

Inflation is the prominent paradigm of the early universe that explains the flatness over cosmological scales, the
 Gaussianity of density perturbations and the near
scale invariance of the cosmic power spectrum. Accelerated cosmic expansion
during inflation pushes perturbation modes from casually connected scales to  outside the horizon.  
After re-entering the horizon these superhorizon modes provide homogeneity over apparently casually disconnected scales,
and give rise to the peaks in the power
spectrum %you mean power spectrum?
of the cosmic microwave background
(CMB) radiation which has been measured with unprecedented precision over a five-year period by
 the Wilkinson Microwave Anisotropy Probe (WMAP5)~\cite{Hinshaw:2008kr}.  
%In the near future 
The Planck~\cite{bib:planck} satellite, planned to be launched in 2009, and continued observation by WMAP 
will further exploit the rich information from both CMB temperature and polarization power spectra.

%In spite of the promising solution to the horizon problems
However, the mechanism that drives the early universe into inflation remains an open question. 
Generically inflation can be modelled by an inflationary field rolling down a potential~\cite{bib:slowroll,Liddle:1994dx}. 
Models may be large field~\cite{bib:lf}, small field~\cite{bib:sf} and hybrid~\cite{Linde:1993cn} and 
have been widely studied.  Alternatively an inverse method~\cite{bib:method} focuses solely on the kinematics of rolling and reconstructs the 
inflationary potential in a model-independent manner. The
slow-roll parameters are defined in terms of the derivatives of the potential.
These parameters can determine the primordial power spectrum
that sheds light on how well the inflationary potential can be experimentally probed. 
Slow-roll parameters have been utilized lately to analyze inflation with WMAP data~\cite{bib:wmapsl} and the upcoming
Planck project~\cite{bib:plancksl,Adshead:2008vn}.

A number of radio telescopes are currently being proposed, planned or constructed to observe the redshifted 21 cm hydrogen line from the 
Epoch of Reionization (EoR), {\it e.g.}, MWA~\cite{bib:wma}, 21CMA~\cite{bib:cma}, LOFAR~\cite{bib:lofar}, 
 GMRT~\cite{bib:gmrt}, PAPER~\cite{bib:paper}, Square Kilometer Array (SKA)~\cite{bib:ska},
and Fast Fourier Transform Telescope (FFTT)~\cite{bib:fftt}.  
21 cm tomography maps the neutral hydrogen in the universe over a wide range of redshifts and provides a promising cosmological probe, 
with arguably greater potential than CMB and galaxy surveys.   
Several studies have investigated the accuracies with which cosmological parameters can be measured by upcoming 21 cm experiments, both by mapping diffuse hydrogen before and during the EoR~\cite{bib:preEoR} and by mapping neutral hydrogen in the galactic halo after reionization~\cite{bib:posEoR}.  
In particular, the FFTT experiment optimized for 21 cm tomography can improve measurement of 
the cosmological parameters to an unprecedented level~\cite{Mao:2008ug}.
Consequently, precision measurements from 21 cm tomography open a new window to constrain inflation in the early universe.  

In this paper, we adopt a model-independent approach and forecast how accurately the shape of the inflationary potential can be reconstructed 
by combining the 21 cm data from FFTT or SKA and the CMB data from Planck. In the next two sections we outline the reconstruction method and 
assumptions about the 21 cm power spectrum. In Sections~\ref{classes} and~\ref{analysis}, we describe the two classes of kinematical models 
and their analysis.
%Section~\ref{analysis} discusses experimental specifications and how we constrain the slow-roll and spectral parameters with 21 cm+CMB data, as well as the priors from WMAP5 and number of e-folds during potential shape reconstruction. 
We display our results in Section~\ref{conclusion}.

\section{Potential reconstruction}

We briefly outline the potential reconstruction method and refer the reader to Refs.~\cite{Powell:2007gu, Lidsey:1995np} for extensive
discussions. 

Consider a flat universe whose energy-momentum tensor is dominated by an inflaton field $\phi$ evolving 
monotonically with time in a potential $V(\phi)$.
With the Hubble parameter $H$ expressed in terms of $\phi$, the equation of motion of $\phi$ and the
Friedmann equation can be written as 
%\be
%H^2=\frac{8\pi}{3 m_{Pl}^2}\left(\frac{1}{2}\dot{\phi}^2+V(\phi)\right),
%\label{eq:Friedman}
%\ee
\be
\dot{\phi}=-\frac{m_{{\rm Pl}}^2}{4\pi}H'(\phi),
\label{eq:monophi}
\ee
and
\be
V(\phi)=\frac{3m_{{\rm Pl}}^2}{8\pi} H^2(\phi)\left[1-\frac{1}{3}\epsilon(\phi)\right]\,,
\ee
where $m_{{\rm Pl}}$ is the Planck mass, primes and overdots 
denote derivatives with respect to $\phi$ and time, respectively, and
\be
\epsilon(\phi) = \frac{m^{2}_{{\rm Pl}}}{4\pi}\left(\frac{H'(\phi)}{H(\phi)}\right)^2\,.
\label{eq:pot}
\ee
Inflation occurs so long as $\epsilon < 1$. 
%The expansion then changes from accelerating to decelerating.
%$^.$ denotes a derivative with respect to physical time. The inflaton $\phi$ 
%equation of motion of the inflaton field 
%is described by the equation of motion,
%\be
%\ddot{\phi}+3H\dot{\phi}+V'=0\,,
%\label{eq:KG}
%\ee
%where $'$ denotes a derivative with respect to $\phi$. 
%Combining Eq~\ref{eq:KG} and the time
%derivative of Eq.~(\ref{eq:Friedman}) yields monotunic behavior of 
%$\phi$ evolves 
%monotonically 
%during inflation as
%\be
%\dot{\phi}=-\frac{m_{Pl}^2}{4\pi}H',
%\label{eq:monophi}
%\ee
%Substitue $\dot{\phi}$ in the Friedman equation~\ref{eq:Friedman} with Eq.~(\ref{eq:monophi}) and 
%and the inflationary potential in terms of the Hubble parameter is,
%\be
%V(\phi)=\frac{3m_{Pl}^2}{8\pi} H^2(\phi)\left[1-\frac{1}{3}\epsilon(\phi)\right].
%\ee
%Assuming
%slow-roll the Hubble parameter can be written as an expansion of a single field $\phi$, where
%the expansion coefficients are tagged as the slow-roll parameters.
%In the equation above $\epsilon$ is the lowest order slow-roll parameter defined as

A series of higher order parameters are obtained by successive differentiation~\cite{Kinney:2002qn}:
\be
\lambda_n =
\left(\frac{m^{2}_{{\rm Pl}}}{4\pi}\right)^n\frac{(H'(\phi))^{n-1} H^{(n+1)}(\phi)}{H^n(\phi)}\,, 
\ee
where $n\ge 1$ and the usual slow-roll parameters are $\eta = \lambda_1$ and $\xi=\lambda_2$. No assumption
of slow-roll is made in the definition of these parameters. If the hierarchy of differential equations
is truncated so that $\lambda_n = 0$ for $n\ge m$, an exact solution for $H(\phi)$ (up to a normalization
factor) can be found~\cite{liddle}. Once $H(\phi)$ is known, the shape of the
potential $V(\phi)$ is determined.  

The evolution of the slow-roll parameters is conveniently expressed as a function of the number of e-folds
before the end of inflation $N$. 
%The number of e-folds $N\sim\ln(a_0/a)$ evolves
%monotonically and serves as an alternative time variable.
%evolution of makes $\phi$ an alternative to time $t$ to describe inflation. Anothter alternative 
%time variable is 
With 
\be
\left(\frac{dN}{d\phi}\right)^2=\frac{4\pi}{m_{{\rm Pl}}^2 \epsilon(\phi)}\,,
\ee
%When $a_0$ in the definition of N-efold is the scale factor at the end of inflation, the positve square root of this
% equation gives the correct $\frac{dN}{d\phi}$, which is also monotunic throughout the inflation. 
%By definition $N$ decreases during inflation and $N'$ takes the negative sign.
the flow of the slow-roll parameters is given by~\cite{Liddle:1994dx}
\bea
\label{eq:odes}
\frac{d\epsilon}{dN} & = & 2\epsilon(\lambda_1 - \epsilon)\,,    \\
%\frac{d\eta}{dN} & = & -\epsilon \eta 	+\xi\\	 
\frac{d\lambda_n}{dN} & = & \left[(n - 1)\lambda_1 -n\epsilon\right] \lambda_n + \lambda_{n+1}\,. 
\label{eq:odes1}
\eea
To solve these equations, we need to specify values of the slow-roll parameters when observable modes left the horizon. We denote these 
by a ``0'' subscript and take $k_0=0.05$ Mpc$^{-1}$ to be the fiducial mode. We set $\phi_0=0$.

The spectral indices and their running that define the commonly used power-law parameterization of the  primordial scalar and tensor power 
spectra~\cite{Peiris:2003ff}
\bea
P_s(k)&=&A_s\left(\frac{k}{k_0}\right)^{n_s-1+\half \alpha \ln{\frac{k}{k_0}}}\,,    \\
P_t(k)&=&A_t\left(\frac{k}{k_0}\right)^{n_t}\,,  
\eea
can be related to the slow-roll parameters.
%The amplitudes of the scalar and tensor modes are normalized as
%\be
%A_s=\frac{H^2}{\pi \m_{Pl}^2 \epsilon}, \hspace{1cm} 
%A_t=\frac{16 H^2}{\pi \m_{Pl}^2},
%\ee
To second order, expressions for parameters that will be relevant to our study, are~\cite{Lidsey:1995np}  
\bea
n_s & = & 1+ 2\eta_0 -4\epsilon_0 -2(1+C)\epsilon_0^2 - \half (3-5C)\epsilon_0 \eta_0+\half (3-C)\xi_0\,,\\
\alpha & = & \frac{dn_s}{d \ln k} = -\frac{1}{1-\epsilon_0} \frac{dn_s}{dN}\bigg|_0\,,  
\label{eq:connect2}
\eea
where $C = 4(\ln 2+\gamma)-5$, with $\gamma \sim 0.577$, and the tensor to scalar ratio $r={A_t}/{A_s}$ is
\be
r = 16\epsilon_0[1+2(-2+\ln 2+\gamma)(\epsilon_0-\eta_0)]\,.
\label{eq:connect1}
\ee
WMAP5 data support a red-tilted ($n_s<1$) spectrum and $r<0.25$~\cite{Hinshaw:2008kr}. 
With $A_s$ fixed by observation, the parameter $r$ determines the tensor amplitude. If $r\gtrsim 0.1$, tensor
modes are detectable by Planck~\cite{bib:0.1r}.  
%In case $r$ is negligibly small, models without tensor modes are favored and the rolling becomes significantly slower. We will re-visit
%the role of $r$ in Section~\ref{classes}.

%%% Yi 09/28
\section{21 cm power spectrum}

%%% Yi 01/18/2009
We briefly describe the essential background of 21cm cosmology in this section, and refer the interested reader to a comprehensive review in Ref.~\cite{Furlanetto:2006jb}.
%%%
The redshifted 21 cm line due to the neutral hydrogen hyperfine transition can be measured in terms of the brightness temperature relative to the CMB temperature~\cite{bib:field1959a},
\be
T_b({\bf x})=\frac{3c^3hA_{10}n_H({\bf x})[T_S({\bf x})-T_{CMB}]}{32\pi k_B \nu_0^2 T_S({\bf x})(1+z)^2 \partial v_{||}/\partial r}\,,
\ee
where $A_{10}$ is the spontaneous decay rate of 21 cm transition, $n_H$ is the number density of the neutral hydrogen gas, 
$T_S$ is the spin temperature and $\partial v_{||}/ \partial r$ is
the physical velocity gradient along the line of sight (with $r$ the comoving distance). 
The temperature fluctuation can be parametrized in terms of the fluctuation in the ionized fraction $\delta_x$, the matter density fluctuation 
$\delta$, and the gradient of peculiar velocity along the line of sight $d v_r/d r$.  During the EoR, the hydrogen gas is
heated well above the CMB temperature~\cite{bib:justifyEoR}, so that  in the approximation $T_s \gg T_{\rm CMB}$, 
%With perturbations $T_b$ can be written near its spatial average $\langle T_b \rangle$ as
\be
T_b= \frac{\langle T_b \rangle}{\langle x_H \rangle} [1-\langle x_i \rangle(1+\delta_x)](1+\delta)\left( 1-\frac{1}{H a}\frac{d v_r}{dr}\right)\,,
\ee
where $x_i=1-x_H$ is the ionized fraction of hydrogen gas and $x_H$ is the fraction of neutral hydrogen.  The total 21 cm power 
spectrum $P_{\Delta T}({\bf k})$ is defined by 
\mbox{$\langle\Delta T_b^*({\bf k})\Delta T_b({\bf k}')\rangle \equiv (2\pi)^3 \delta^3({\bf k} - {\bf k'}) P_{\Delta T}({\bf k})$}, 
where $\Delta T_b({\bf k})$ is the deviation from the mean brightness temperature and ${\bf k}$ is the comoving wave-vector that is 
the Fourier dual of the real coordinate position ${\bf r}$.  
We restrict our considerations to
linear perturbation theory ($\delta\ll 1$) and write the Fourier transformed spectrum to leading order as
%\bea
%P_{\Delta T}({\bf k}) & = & \tilde{T_b}^2 \lbrace[\avexh^2P_{\delta \delta}-2\avexh P_{x \delta}+P_{xx}] 	 \\	 
%& & +2\mu^2[\avexh^2P_{\delta \delta}-\avexh P_{x\delta}]+\mu^4 \avexh^2 P_{\delta \delta} \rbrace.	 
%\eea
%where $\mu={\bf k} \cdot \hat{\bf n}$ is the cosine of angle between the wave-vector and the line of sight. $\tilde{T_b}$ is defined as $\tilde{T_b}\equiv \frac{\aveTS-T_{CMB}}{\aveTS} \frac{\aveTb}{\avexh} $. We assume $T_S \gg T_{CMB}$ during
%the epoch of reionization such that the CMB temperature drops out~\cite{bib:justifyEoR}. 
%The angular dependence of the spectrum is in
%the form
\begin{equation}\label{eqn:bright_temp_PS}
P_{\Delta T}({\bf k})=P_0(k)+P_{2}(k) \mu^2 +P_{4}(k) \mu^4\,,
\ee
where the multipole coefficients can be written as
\bea
P_0	& = &	{\cal P}_{\delta \delta} - 2 {\cal P}_{x \delta} + {\cal P}_{x x}\,,    \\
P_2	& = &	2({\cal P}_{\delta \delta} -  {\cal P}_{x \delta})\,,  	\\  
P_4	& = &	{\cal P}_{\delta \delta}\,.
\eea
Here $\mu=\hat{\bf k} \cdot \hat{\bf n}$ is the cosine of angle between the wave-vector and the line of sight.  
The power spectra of matter and 
ionization fluctuations are denoted by ${\cal P}_{\delta \delta}=\tilde{T}^2_b \avexh^2 P_{\delta \delta}$, 
${\cal P}_{x \delta}=\tilde{T}^2_b \avexh \avexi P_{\delta_x \delta}$, and 
${\cal P}_{x x}=\tilde{T}^2_b \avexi^2 P_{\delta_x \delta_x}$, where 
$\tilde{T_b}\equiv \frac{\aveTS} {\aveTS-T_{CMB}}\frac{\aveTb}{\avexh} \approx \frac{\aveTb}{\avexh}$. 
%We assume that
%adopt the ``optimistic'' assumption ~\cite{Mao:2008ug} that 
%at the start of the EoR, the fluctuation in ionized fraction is negligible compared to the matter density fluctuation so that
% ${\cal P}_{x \delta} = {\cal P}_{x x} \approx 0 $. Under this optimistic assumption $P_{\Delta T}$ depends only on
%the matter power spectrum that depends only on cosmology. Note however, that the results of Ref.~\cite{Mao:2008ug} show that
%inclusion of the ionization power spectra degrades the precision of parameter estimation only slightly for a large dataset.
%%% Added by Yi 10/17/2008
We account for ionization effects by parameterizing the ionization power spectra as~\cite{Mao:2008ug} 
\begin{eqnarray}
\label{reion1}
{\cal P}_{x x}(k) & = & b^2_{xx} \left[ 1+\alpha_{xx}(k\,R_{xx}) + \,(k\,R_{xx})^2\right]^{- {\gamma_{xx} \over 2}} {\cal P}_{\delta \delta}\,,\\ 
{\cal P}_{x \delta}(k) & = & b^2_{x\delta} \,\exp{\left[-\alpha_{x\delta} (k\,R_{x\delta})-(k\,R_{x\delta})^2\right]}  {\cal P}_{\delta \delta}\,,
\label{reion2}
\end{eqnarray}
where $b^2_{xx}$ and $b^2_{x\delta}$ are the amplitudes of the spectra, $R_{xx}$ and $R_{x\delta}$ are the effective 
sizes of the ionized bubbles (HII regions), and $\alpha_{xx}$,  $\gamma_{xx}$ and $\alpha_{x\delta}$ are spectral indices.
We adopt the fiducial values of Table III in Ref.~\cite{Mao:2008ug}.  
%Here ${\cal P}_{\delta \delta}^{\rm (fid)}$ is the fiducial matter power spectrum.    
%Using the total power spectrum from 21 cm measurements, the cosmological parameters can be constrained during the 
%EoR by marginalizing over the ionization parameters.  

\section{Model classification}\label{classes}

Kinematically different potentials can be categorized 
based on the relative sizes of slow-roll parameters. The parameter 
$\epsilon$ plays a critical role that determines the duration of inflation, the rate of change of $\phi$,
how much the inflationary potential $V(\phi)$ rolls down from its initial height, and the tensor to scalar ratio. 
We follow a recent classification that is based on the size of $\epsilon$~\cite{Adshead:2008vn}. 
%To the leading order $\epsilon$ determines the tensor to scalar ratio, hence the fast/slow initial rolling also physically represent significant/low gravitational wave presence.
%As the slow-roll parameters higher than the 2nd order
%are not effectively constrained by the spectral parameters we take the following 'models' 
%in this analysis:

\newpage
\subsection{High $\epsilon$ models.} 

High $\epsilon$ models yield $r \gtrsim 0.1$ so that tensor modes are detectable by Planck. 
%A sizeable $\epsilon$ dominates the rolling speed during and the duration of inflation. 

{\bf One-parameter models.} $\epsilon$ is the sole parameter in these models and determines the primordial
spectra. As the only free parameter, $\epsilon$ is stringently constrained by 21 cm and CMB data. However, this single parameter scenario is 
not easily realized in particle physics.

{\bf Two-parameter models.} In these models $\eta$ contributes to the evolution equations. 
Two-parameter models resemble a $\Lambda$CDM cosmology with significant tensor power.

{\bf Three-parameter models.} Inflationary rolling is described by $\epsilon$, $\eta$ and $\xi$. These models resemble
a $\Lambda$CDM model with measurable $r$ and a large $\xi$ can contribute significantly to the running of scalar spectral index $\alpha$, breaking 
scale invariance of the power spectrum. The non-zero $\xi$  allows the rolling to speed up at late times and gives a variety of shapes for
 the potential. $\xi$ contributes significantly to $\alpha$ when it is numerically comparable to the other two parameters. 
Generically, $\xi$ speeds up the evolution of $\epsilon$ and 
 a large $\xi$ causes a prompt end to inflation with small $N$. 

%\vspace{0.5cm}
%In all the three high $\epsilon$ models, $\alpha$ is slightly negative but consistent with $\alpha$=0 within the 2$\sigma$ 21 cm+Planck range. 

\subsection{Low $\epsilon$ models.} 

In these models $\epsilon$ is vanishingly small when $k_0$ leaves the horizon. We set $\epsilon_0=10^{-8}$.
This represents extremely slow rolling at horizon-crossing.
In such models non-zero higher order parameters cause $\epsilon$ to grow super-exponentially near the end of inflation and the potential falls
abruptly with a cliff-like feature. 

{\bf Two-parameter models.} These models resemble $\Lambda$CDM with near scale invariance in the power spectrum and negligible tensors. 
The parameter $\eta$ can be strongly constrained but the number of e-folds are generally large because an efficient accelerating mechanism is absent. 
Within 95\% C.~L. constraints from WMAP5, we find that these models give $N>180$. A large $N$ indicates that inflation must end via a hybrid transition. 

{\bf Three-parameter models.} A non-zero $\xi$ parameter speeds up rolling, significantly lowers the number of e-folds and allows a 
non-zero $\alpha$. These models can easily be distinguished from the two-parameter case. 
It is noteworthy that in these models it is possible for rolling to be even slower than in two-parameter models during most of the inflationary 
period. This is followed by significant late-time acceleration which causes the overall effect of $\xi$ to be a speed-up of rolling.
The phase of slow evolution also occurs in models with higher order kinematical parameters.

Here we do not investigate low $\epsilon$ models with higher order
parameters ($\lambda_n$ with $n\ge 3$) since such models are 
indistinguishable from the three-parameter model.
%and large allowed region make such models difficult to be distinguished from 
%the two parameter model 
%and these higher order slow-roll parameters lack effective cosmological constraints as well.

%%% Yi 09/28
\section{Analysis}\label{analysis}

21 cm experiments do not directly measure ${\bf k}$ or $P_{\Delta T}({\bf k})$. The power spectrum $P_{\Delta T}({\bf u})$ is evaluated 
in the observer's 
pixel ${\bf u}$ that is the Fourier dual of the observed vector \mbox{${\bf \Theta} \equiv \theta_x \hat{x} + \theta_y \hat{y} + \Delta \nu \hat{z}$} 
where $(\theta_x,\theta_y)$ gives the angular location on the sky plane,  $\Delta \nu$ is the frequency difference from the central redshift of a 
data bin and the $z$-axis is along the line of sight.  By using $P_{\Delta T}({\bf u})$ instead of $P_{\Delta T}({\bf k})$, we avoid
the Alcock-Paczynski effect~\cite{bib:AP}, which arises from the model dependence in the projection of the physical wave-vector ${\bf k}$ over 
 cosmological distances.

We employ the Fisher matrix formalism to determine the precision of parameter estimation. 
%%Oct 4
Following Ref.~\cite{Mao:2008ug},
we resolve the 21 cm spectrum $P_{\Delta T}({\bf u})$ 
into pixels and the 21 cm Fisher matrix is constructed as
\be
{\bf F}^{21 cm}_{ab}=\sum_{pixels}\frac{1}{[\delta P_{\Delta T}({\bf u})]^2}
\left( \frac{\partial P_{\Delta T}({\bf u})}{\partial \lambda_a} \right)
\left( \frac{\partial P_{\Delta T}({\bf u})}{\partial \lambda_b} \right)\,,
\label{eq:fm21cm}
\ee
where $\delta P_{\Delta T}({\bf u})$ is the power spectrum measurement error in a pixel at ${\bf u}$ and ${\boldsymbol \lambda}$ is 
%%%%%By Yi Oct17
the combined set of cosmological and ionization parameters.

%%%%%By Yi Oct7
We consider 21 cm measurements in the redshift range $6.8-8.2$ with three redshift bins centered at $z=7.0$, 7.5 and 8.0, with a nonlinear 
cut-off scale 
$k_{max}=2$ Mpc$^{-1}$, and 16000 observation hours.  Non-Gaussianity of ionization signals is assumed to be negligible in our analysis.  
%%% Yi 01/18/2009
We assume that the foreground can be perfectly cleaned above the scale $ k_{\rm min} = 2\pi/yB $ where $yB$ is the comoving line-of-sight distance width of a single redshift bin.  This assumption was shown to be a good approximation in Ref.~\cite{bib:preEoR}. 
%%%
We consider two detector arrays, SKA and FFTT, which have optimal signal-to-noise ratios among planned 21 cm experiments.  
We assume an azimuthally symmetric distribution of baselines in both arrays.  
The design of SKA has not been finalized.  We adopt the ``smaller antennae'' version of SKA, in which the array will have 7000 10~m antennae. 
We assume that 16\% of the antennae are concentrated in a nucleus within which the area coverage fraction is close to 100\%; 4\% of the antennae 
have a coverage density that falls as the inverse square of the radius; and 30\% are in the annulus where the coverage density is low but rather uniform
out to a 5~km radius. 
We ignore the measurements from the sparse distribution of the remaining 50\% of the antenna panels that are outside the annulus.  
FFTT is a future square kilometer array optimized for 21 cm tomography as described in Ref.~\cite{bib:fftt}.
Unlike other interferometers, which add in phase the dipoles in each panel or station, FFTT can obtain more information by correlating all of its dipoles.  We assume that FFTT contains
a million 1~m $\times$ 1~m dipole antennae in a contiguous
core subtending a square kilometer, and providing a field-of-view of
$2\pi$ steradians.  
%For definitions of our ``optimistic'' assumptions for the 21 cm power spectrum and experimental specifications, see Ref.~\cite{Mao:2008ug}.
%%%%

The Fisher matrix formalism for the CMB is well established~\cite{Tegmark:1996bz};
for Planck data we follow the latest experimental specifications~\cite{bib:planck}.
We include both temperature and polarization
 measurements and assume $l_{max}=3000$ with three frequency channels while the other channels are used 
for foreground subtraction. 
The CMB power spectra's parameter dependence is computed
using the Code for Anisotropies in the Microwave Background (CAMB)~\cite{bib:camb}. 

The Fisher matrix is cosmology dependent and we work in the flat ($\Omega_k=0$) standard $\Lambda$CDM 
model and fix  $\Omega_\nu h^2=0.0074$ (neutrino density) and $Y_p=0.24$ (helium abundance).
The fiducial values of the non-slow-roll parameters are set near the best-fit of the
 WMAP5 result~\cite{Hinshaw:2008kr}: $h=0.72$ (Hubble parameter $H_0 \equiv 100h$ km s$^{-1}$ Mpc$^{-1}$), $\tau=0.087$ 
(reionization optical depth), $\Omega_\Lambda=0.742$ (dark energy density), 
$\Omega_b h^2=0.02273$ (physical baryon density), and $A_s=0.9$. We fix 
${\cal P}_{\delta \delta} (k)$ in Eqs.~(\ref{reion1}) and~(\ref{reion2}) when varying cosmological
parameters, so that constraints arise only from the ${\cal P}_{\delta \delta}$ terms in $P_0$,
$P_2$ and $P_4$.

The Fisher matrices depend on
${\boldsymbol \lambda}$ that includes ($n_s$, $r$, $\alpha$) in  ${\bf F}^{Planck}$ and ($n_s$, $\alpha$) in  ${\bf F}^{21cm}$ in addition 
to the non-inflationary parameters. We marginalize over the latter to obtain ${\bf F}_{(n_s, r, \alpha)}^{Planck}$ and 
${\bf F}_{(n_s, \alpha)}^{21cm}$.
The Jacobian matrix $\partial {\boldsymbol \lambda_{spec}}/\partial {\boldsymbol \lambda}_{sr}$ (where the subscript ``spec'' indicates ($n_s$, $r$, $\alpha$) for Planck and ($n_s$, $\alpha$) for 21 cm experiments),
can be used to obtain the Fisher matrices for
the slow-roll parameter set ${\boldsymbol \lambda}_{sr}\equiv(\epsilon, \eta, \xi)$,
\be
{\bf F}_{sr}= \bigg(\frac{\partial {\boldsymbol \lambda_{spec}}}{\partial {\boldsymbol \lambda}_{sr}}\bigg)^T {\bf F}_{spec} \, \frac{\partial {\boldsymbol \lambda_{spec}}}{\partial {\boldsymbol \lambda}_{sr}}\,.
\ee
The three independent 
spectral parameters allow the Jacobian matrix
a maximal rank of three, and directly constrain up to three slow-roll parameters. 
%Then the Fisher matrix of slow-roll parameters $\tilde{\bf F}_{sr}$ is obtained by marginalizing over ${\boldsymbol \lambda}_c$ in $\tilde{\bf F}$.
We consider Planck and 
21 cm data independently, so the combined Fisher matrix is the sum of the contributions, 
\be
{\bf F}_{sr}^{tot} = {\bf F}_{sr}^{21cm} + {\bf F}_{sr}^{Planck}\,,
\ee
%where the ``sr'' subscript denotes that ${\bf F}_{sr}$ contains only the slow-roll parameters.
%A Jacobian matrix transforms the Fisher matrix on to
%a set of parameters {$\tilde{\boldsymbol \lambda}$} that contains the slow-roll and non-spectral parameters, 
%\be
%\tilde{{\bf F}}_{a' b'}= \sum_{ab} \frac{\partial \lambda_a}{\partial \tilde{\lambda}_{a'}} {\bf F}_{ab} \frac{\partial \lambda_b}{\partial \tilde{\lambda}_{b'}},
%\ee
%where ${\partial {\boldsymbol \lambda}}/{\partial \tilde{\boldsymbol \lambda}}$ changes the parameter set ($n_s$, $r$, $\alpha$ ,${\boldsymbol \lambda}_c$) into (${\boldsymbol \lambda}_{sr}$, ${\boldsymbol \lambda}_c$)
%according to Eqs.~(\ref{eq:connect1}) and~(\ref{eq:connect2}). ${\boldsymbol \lambda}_{sr}$ and ${\boldsymbol \lambda}_c$ denote the slow roll parameters $\{\epsilon,\eta,\xi\}$ and non-spectral parameters, respectively.
which we use to construct a $\chi^2$ function, 
\be
\chi^2({\boldsymbol \lambda}_{sr})= {{\boldsymbol \delta}^T_{\lambda_{sr}}} {\bf F}_{sr}^{tot} {\boldsymbol \delta}_{\lambda_{sr}}\,,
\ee
where ${\boldsymbol \delta}$ denotes the deviations from the fiducial values of the slow-roll parameters. 
%$\tilde{{\bf V}}_{sr}^{tot}=\left({\tilde{{\bf F}}^{tot}_{sr}}\right)^{-1}$ is the slow-roll covariance matrix with the non-spectral cosmological parameters ﻿marginalized over.%$\tilde{{\bf V}}_{sr}$ is the block that contain the errors of ${\boldsymbol \lambda}_{sr}$ inside the full covariance matrix $\tilde{{\bf V}}$=${\tilde{{\bf F}}}^{-1}$. Nonslow-roll parameters are marginalized over.

%Since the 21 cm data do not constrain $r$, care is required when all three slow-roll parameters ($\epsilon, \eta, \xi$) are simultaneously 
%included in the analysis. In the three-parameter case the Jacobian matrices are multiplied with the Fisher matrices 
%${\bf F}_{(n_s, \alpha)}^{21cm}$ and ${\bf F}_{(n_s, r, \alpha)}^{Planck}$ after marginalizing over ${\boldsymbol \lambda}_c$.

\section{Results}\label{conclusion}

\begin{table*}
\begin{center}
\scriptsize
\begin{tabular}{r|c|c|c|c}
\hline
 Model\hspace{0.55cm}	& Fiducial & 1$\sigma$ (Planck alone) & 1$\sigma$ (SKA+Planck)	& 1$\sigma$ (FFTT+Planck)\\
\hline
\multicolumn{1}{l}{\hspace{1.15cm} High $\epsilon$}\\
\hline
%Exx	
1 parameter,\hspace{0.2cm}   $\epsilon$ \hspace{0.4cm}	& 0.0071	& 6.9$\times 10^{-4}$ &	6.3$\times 10^{-4}$	&	6.9$\times 10^{-5}$		\\
\hline
%EEx				
2 parameter,\hspace{0.2cm} $\epsilon$	\hspace{0.4cm} & 0.0053	&0.0014&	0.0014	&	0.0012			\\
$\eta$	\hspace{0.4cm} &	-0.013 &	0.0034 &	0.0033	&	0.0026			\\
\hline
%EEX
3 parameter,\hspace{0.2cm} $\epsilon$ \hspace{0.4cm}	& 0.0063	& 0.0015	&	0.0015	&		0.0014		\\
$\eta$ \hspace{0.4cm}	&	0.0069 & 0.0036	&	0.0033	&		0.0028		\\
$\xi$ \hspace{0.4cm}	& 0.00083	& 0.0026	&	0.0016	&		1.6$\times 10^{-4}$		\\
\hline
\multicolumn{1}{l}{\hspace{1.15cm} Low $\epsilon$}\\
\hline
%xEx
2 parameter,\hspace{0.2cm}  $\epsilon$	\hspace{0.4cm} &	10$^{-8}$ & ---	&	---	&		---		\\
$\eta$	\hspace{0.4cm} & -0.027	& 0.0016 &	0.0014	&	1.5$\times 10^{-4}$	\\
\hline
%xEX
3 parameter,\hspace{0.2cm} $\epsilon$	\hspace{0.4cm} & 10$^{-8}$	& ---	&	---	&		---		\\
$\eta$	\hspace{0.4cm} & -0.0069 &	0.0024 &	0.0016	&	2.2$\times 10^{-4}$			\\
$\xi$	\hspace{0.4cm} & 0.002	& 0.0026	&	0.0016	&		1.4$\times 10^{-4}$		\\
\hline
\end{tabular}
\normalsize
\caption{\footnotesize Uncertainties on slow-roll parameters for models classified according to the size of $\epsilon$. 
 The fiducial values at the time of horizon-crossing are chosen to be consistent with the $2\sigma$ ranges favored by WMAP5 data~\cite{Hinshaw:2008kr}.
% The fiducial parameter values are chosen such that different models occupy separate regions in the spectral parameter space and the highest order slow-roll parameter does not reach zero in the 21 cm+Planck analyses. However, non-zero high order slow-roll parameters lead to faster rolling and less e-folds. In the high $\epsilon$ three-parameter case the 2$\sigma$ SKA+Placnk uncertainty of $\xi$ exceeds the largest $\xi$ value that satifies the e-folds prior $N>$30.
%The 21 cm+Planck constraint on 
%cosmological parameters and reconstructed potentials are shown in Fig.~\ref{fig:fid_pot} and Fig.~\ref{fig:cos_comp}. 
}
\label{tab:category}
\end{center}
\end{table*}

\begin{figure}[ht]
\begin{center}
\includegraphics[scale=0.7]{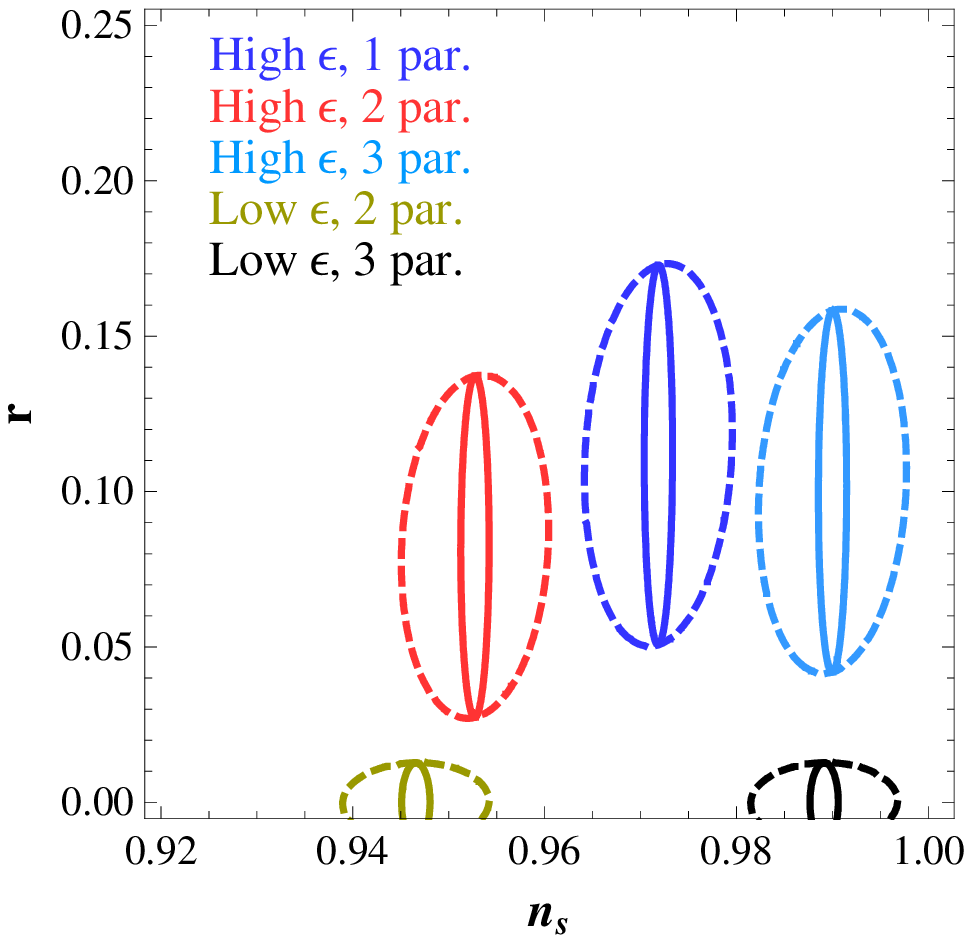}
\hspace{0.6cm}
 \includegraphics[scale=0.7]{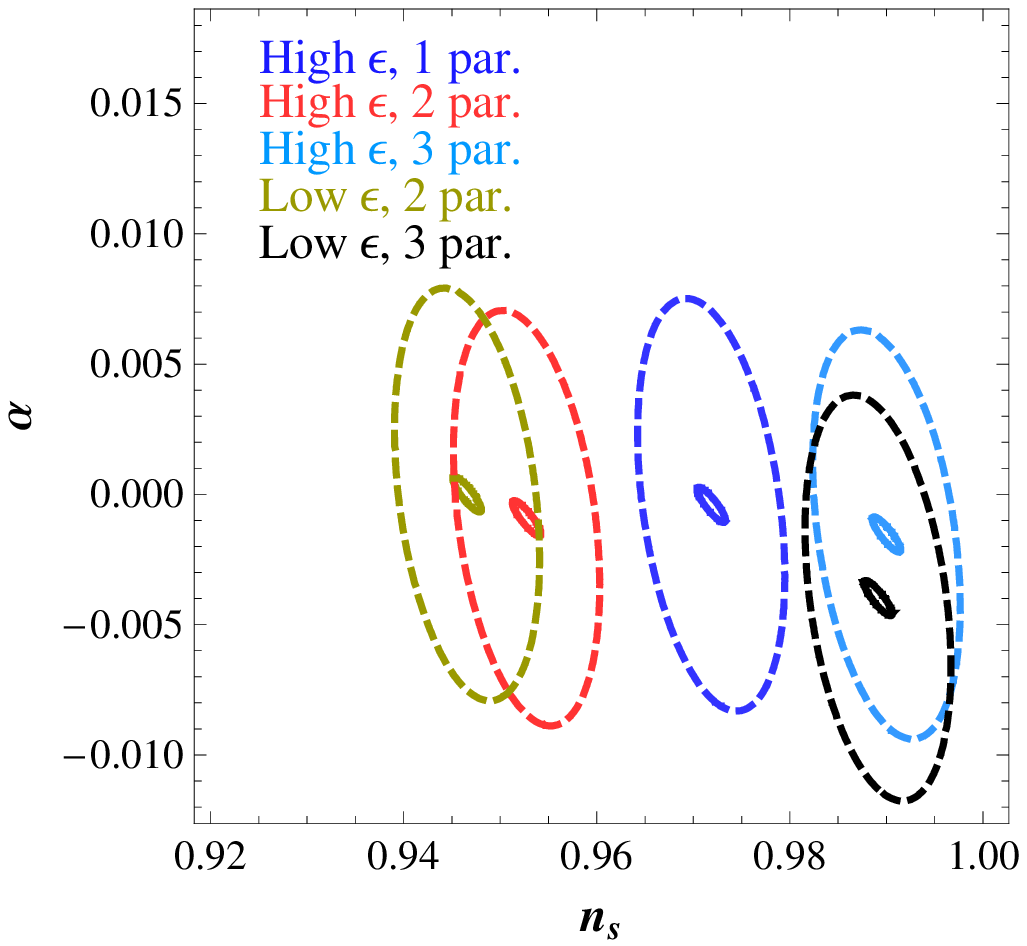}
\caption
{\footnotesize  2$\sigma$ forecasts for the fiducial points in Table~\ref{tab:category} from a Fisher matrix  analysis of SKA+Planck (dashed) 
and FFTT+Planck (solid) in the ($n_s, r$) and ($n_s, \alpha$) planes. 
%The allowed regions for FFTT+Planck appear as thick dots in the  ($n_s, \alpha$) plane because of their small size.
%In the labels ``n par.'' stands for the number n of slow-roll parameters in each model.
}
\label{fig:fid_pot}
\end{center}
\end{figure}

\begin{figure}[ht]
\begin{center}
  \includegraphics[scale=0.7]{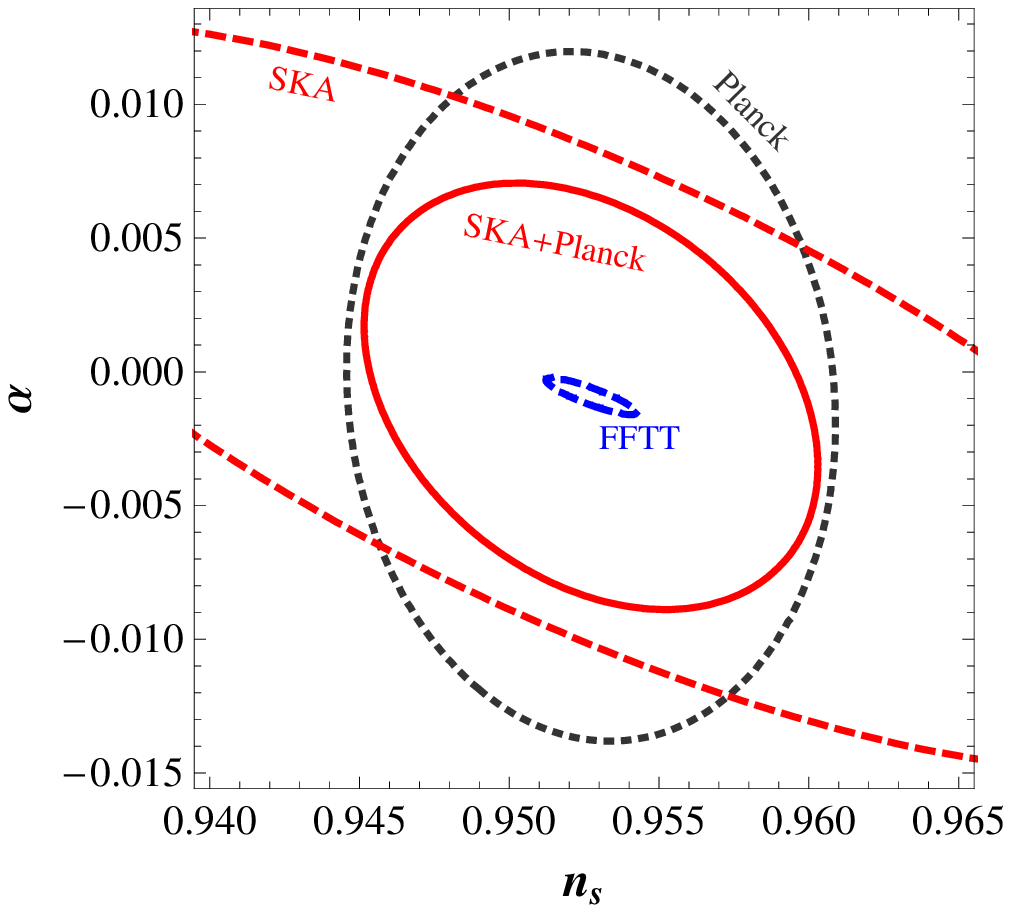}
\caption
{\footnotesize  The impact of 21 cm experiments on parameter estimation. $2\sigma$ regions from an analysis of Planck alone, 21 cm alone, 
and 21 cm+Planck. The fiducial point for the high $\epsilon$ two-parameter model is used. 
FFTT and Planck are complementary: FFTT has good sensitivity to $\alpha$ but no sensitivity to $r$ and Planck
has good sensitivity to $r$ but not $\alpha$. 
We do not show the FFTT+Planck ellipse since it is indistinguishable from the ellipse for FFTT alone.
% in the ($n_s, r$) plane the Planck alone ellipse (dotted gray) also represents the allowed region of the SKA+Planck analysis.
}
\label{fig:cos_comp}
\end{center}
\end{figure}

We forecast constraints on the slow-roll
parameters at the fiducial points of Table~\ref{tab:category} that are consistent with WMAP5 results. 
%Table~\ref{tab:category} shows for each model (discussed later in Section V) the joint 21 cm+Planck analysis puts more stringent constraints on the 
%slow-roll parameters than Planck data alone. 
To supplement the uncertainties listed in the table, we provide the corresponding (approximate) uncertainties for the more familiar spectral parameters. 
The joint SKA+Planck (FFTT+Planck) analysis gives the 1$\sigma$ uncertainties $\delta n_s=0.0031$, 
$\delta \alpha=0.0032$ (\mbox{$\delta n_s=6\times 10^{-4}$}, $\delta \alpha=2.7\times 10^{-4}$).  
%%% Yi 01/18/2009
These results roughly apply to all the classes of models in Section~\ref{classes}.
%%%
These uncertainties are larger, but consistent with those in Ref.~\cite{Mao:2008ug} since we marginalize over 
all other parameters, while in Ref.~\cite{Mao:2008ug}, 
$r$ and $\alpha$ are held fixed in computing uncertainties for $n_s$, and $r$ is fixed in computing uncertainties for $\alpha$. 
For $r$ large enough to be measured by Planck,
$\delta r|_{r\sim 0.1}=0.022$ and if $r$ is tiny, a bound $\delta r|_{r\sim 0}=0.005$ can be placed at $1\sigma$. 21 cm data do not add 
any information on tensor modes. 
Figure~\ref{fig:fid_pot} shows this information pictorially. The fiducial points are chosen so that the allowed regions have
minimal overlap. A comparison of the constraints from the joint analyses with that from Planck data alone
is shown in Fig.~\ref{fig:cos_comp}. The constraining power of 21 cm data comes from their sensitivity to $n_s$ and particularly $\alpha$. 
21 cm and Planck data are complementarity in their sensitivity to $\alpha$ and $r$.

%%% Yi 01/18/2009
It should be mentioned that higher order corrections to the brightness temperature power spectrum 
(Eqs.~\ref{eqn:bright_temp_PS} -- \ref{reion2}) may lead to errors as large as $\mathcal{O}(1)$ in the power spectrum at small scales $k \gtrsim 1\, h  \,{\rm Mpc}^{-1}$ when the neutral fraction is $ \left< x_H \right> \sim 0.5$~\cite{Lidz:2006vj}.   
However, since interferometer array measurements are more sensitive to small $k$ modes than to large 
$k$ modes because of thermal noise, cosmological constraints depend only weakly on $k_{\rm max}$, the nonlinear cutoff scale above which we ignore 21 cm contributions to cosmology.  Figure 6 of 
Ref.~\cite{Mao:2008ug} shows that in their setup, the uncertainty in the tilt measured by the FFTT and the Planck data varies from roughly 0.0003 to 0.0006 to 0.0009 as $k_{\rm max}$ is reduced from 
2~${\rm Mpc}^{-1}$ to~1 ${\rm Mpc}^{-1}$ to 0.6~${\rm Mpc}^{-1}$.  Regardless of the exact value of $k_{\rm max}$ that can be determined by further careful modeling,  
it is qualitatively robust that cosmological constraints from FFTT and Planck data will reach  unprecedented precision, {\it e.g.}, the measurement of $n_s$ at the level of $10^{-4}$.  

\begin{figure}[ht]
\begin{center}
 \includegraphics[scale=0.70]{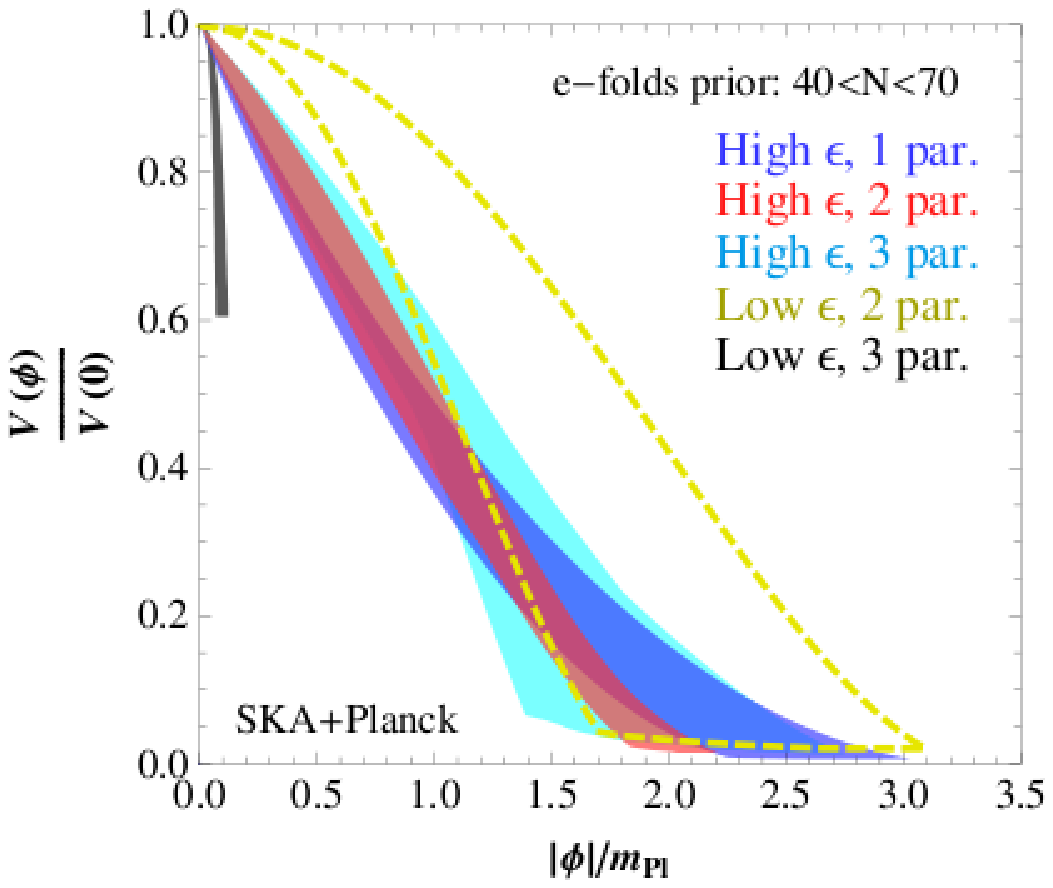} \hspace{1cm}
 \includegraphics[scale=0.70]{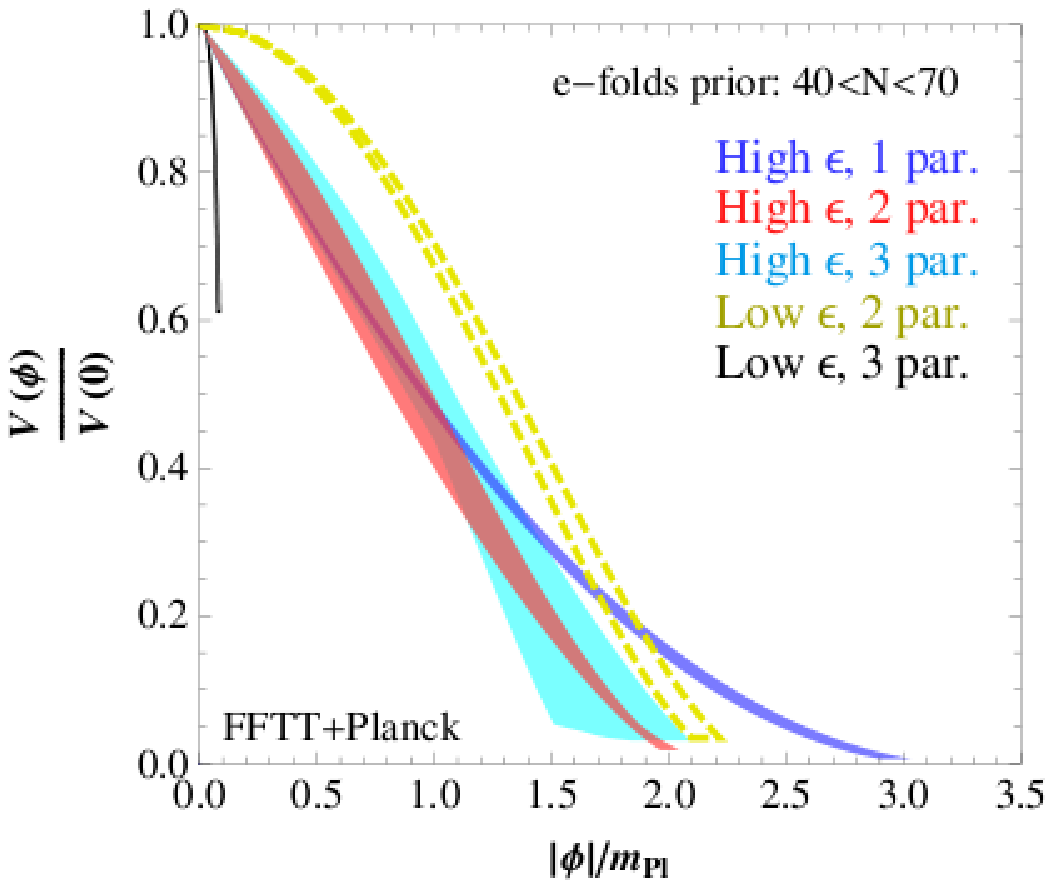}\\
 \includegraphics[scale=0.70]{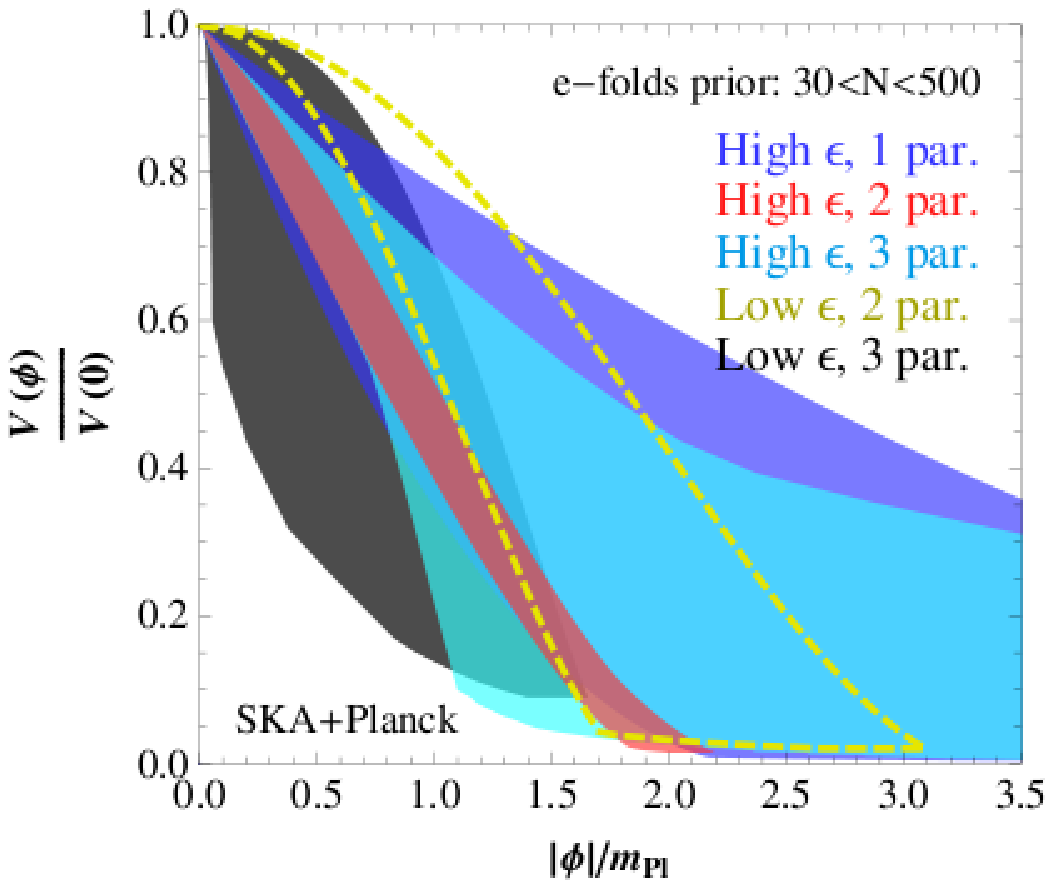} \hspace{1cm}
 \includegraphics[scale=0.70]{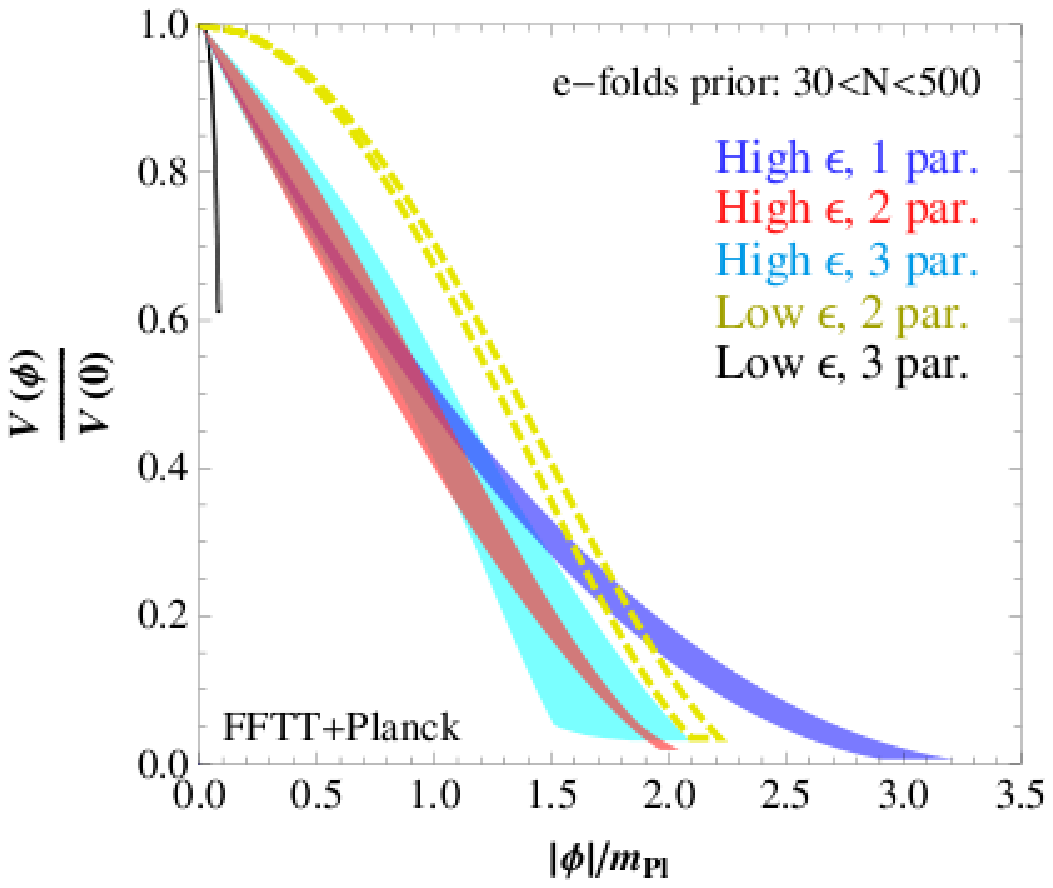}
\caption{\footnotesize
%potentials at the fiducial point in each model (left) and the 
Bands of reconstructed
potentials at $2\sigma$ from SKA+Planck (left) and FFTT+Planck (right) for two sets of priors on the number of e-folds, $40<N<70$ (upper) 
and $30<N<500$ (lower). Note that the low $\epsilon$ two-parameter model requires $N>180$ and is eliminated by the $40<N<70$ prior. The unshaded bands in the upper panels are shown only for
comparison.  It is remarkable that the reconstruction using FFTT+Planck is barely affected by the e-fold prior. 
The enlarged SKA+Planck band for the low $\epsilon$ three-parameter model for $30<N<500$ is a consequence of
$\xi\simeq 0$ being allowed at $2\sigma$. 
%and the priors on the initial values of slow-roll parameters include (0.92$<n_s<$0.99, $r<$0.25).
% The fiducial points are chosen such that
%the highest-order parameter in each model remains non-zero inside the 2$\sigma$ range, which allows models to be
% more visually separated.  However it should be noted that with close fiducial values 
% an n-parameter model can cover up the range of the (n-1)-parameter model when the nth-order parameter 
%reaches zero. 
%In the high $\epsilon$ models the relatively faster rolling speed makes
%the allowed bands sensitive to the lower limit of $N$. The 40$<N<$500 prior cuts off a
% number potentials with $N<$40 and the allowed bands for the high $\epsilon$ one and two parameters are even narrower 
% than those with the 30$<N<$80 prior. 
%In the low $\epsilon$ three-parameter model, late time acceleration suppresses rolling speed causes limited growth of $\Delta\phi$, $\Delta V$. 
%Note that the low $\epsilon$ two-parameter model is eliminated by the 30$<N<$80 prior. This model has a range of number of e-folds $N>$180 within the WMAP5 prior and is shown as unfilled bands. 
}
\label{fig:bands}
\end{center}
\end{figure}

To implement Monte Carlo reconstruction of the potential,  we randomize slow-roll parameters inside the $2\sigma$ regions allowed by 
21 cm+Planck data as the values when the scale $k_0$ left the horizon. We then evolve Eqs.~(\ref{eq:odes}) and~(\ref{eq:odes1}) forward in time.  
Those cases are selected in which inflation ends with the number of e-folds $N$
that pass a prior $N_{min}<N<N_{max}$. The prior on $N$ is necessary because (i) a sufficiently 
large $N$ is required to be consistent with the observed horizon size; (ii) a small $N$ indicates relatively fast rolling
which suggests that higher-order parameters may not be small enough to be truncated; (iii) 
a large $N$ indicates that rolling is extremely slow so that a hybrid mechanism might be responsible for end the inflation.
While our framework supposes that observable inflation is dominated by a single scalar field, it does not preclude the
possibility of a hybrid transition caused by other fields ending inflation.
We use two priors, $40<N<70$ and $30<N<500$. The first prior is typical for a plausible expansion history of our universe with Ref.~\cite{leach} arguing
for $N$ between 50 and 60. This first prior does not account 
for a hybrid transition. Our second prior is rather conservative $30<N<500$, with the large values suggesting that some other mechanism brings an
abrupt end to inflation.

\begin{figure}[ht]
\begin{center}
 \includegraphics[scale=0.7]{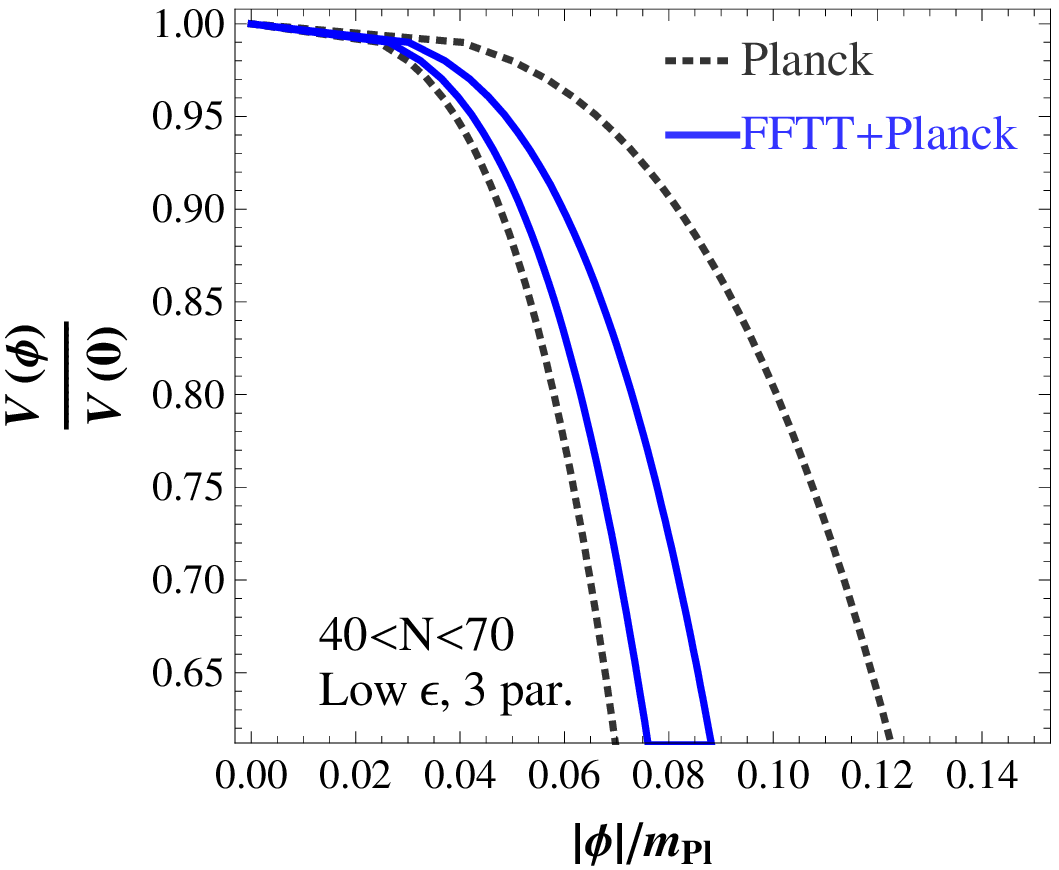}
 \includegraphics[scale=0.7]{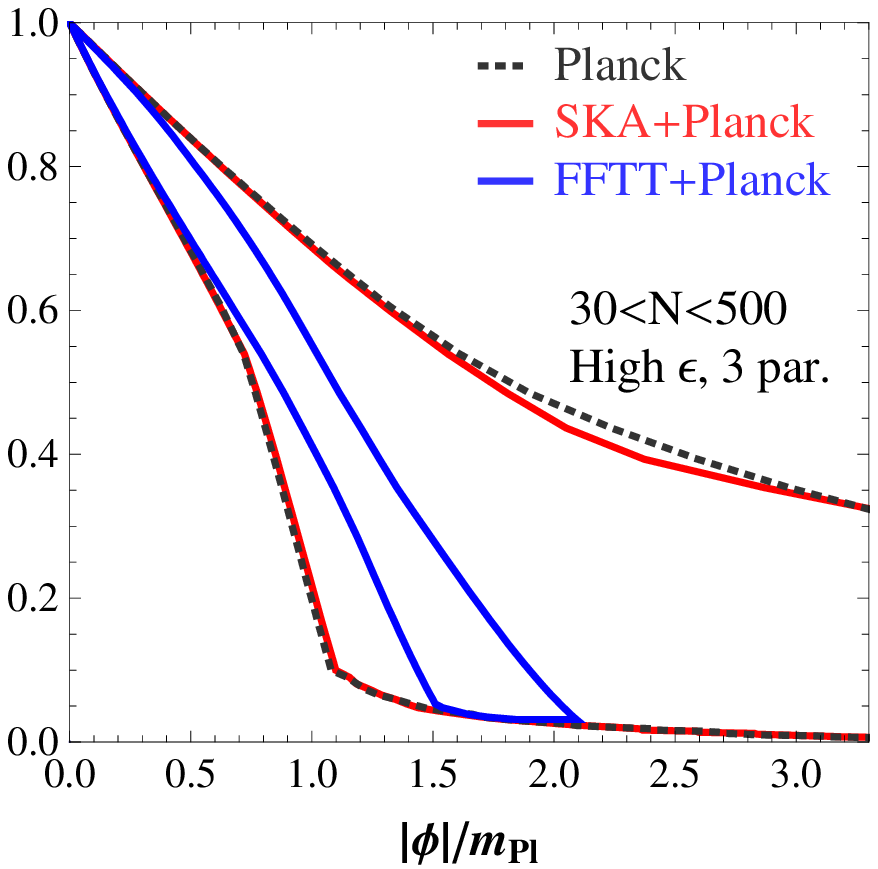}
\caption
{\footnotesize The impact of 21 cm experiments on potential reconstruction. $2\sigma$ bands from Planck alone and 
21~cm+Planck. The left panel is a magnified view of the low $\epsilon$ three-parameter model; 
the SKA+Planck band is almost identical to that for Planck alone and is not shown.
}
\label{fig:bands_comp}
\end{center}
\end{figure}

In Fig.~\ref{fig:bands}, the bands show the envelopes of possible potentials at
the $2\sigma$ C.~L. for each class of models with fiducial values as in 
Table~\ref{tab:category}. The envelopes capture the shapes of the potentials because
the reconstructed potentials do not show any fine dependence on $\phi$.
While the reconstruction from SKA+Planck is clearly affected by the e-fold prior, it is 
striking that the reconstruction from FFTT+Planck is essentially unaffected. 
The low $\epsilon$ two-parameter model is inconsistent with the $40<N<70$ prior since WMAP5 data yield $N>180$ for
these models.
The SKA+Planck band for the low $\epsilon$ three-parameter model expands greatly for the $30<N<500$ prior because
$\xi\simeq 0$ becomes allowed at $2\sigma$. 
Note that detection of tensors by Planck is not sufficient 
to guarantee satisfactory potential reconstruction using Planck data alone. For example, FFTT data crucially improve the 
reconstruction of the high $\epsilon$ 1 parameter model.
In Fig.~\ref{fig:bands_comp}, we compare results for models which require 3 slow-roll parameters for their 
description. FFTT data narrow down the 2$\sigma$ bands considerably. 
%The bands from SKA+Planck data are almost identical to those from Planck data alone, and are not shown.

We conclude by emphasizing that a
%current 21 cm projects target the redshift range of $7\lesssim z \lesssim 12$ or even higher.
 joint analysis of 21 cm measurements from FFTT with Planck data will significantly
help pin down the slow-roll parameters and determine the shape of the inflationary potential. The improvement
over the reconstruction using Planck data alone may stimulate major developments in our understanding of the
particle physics responsible for inflation.

%\newpage
\section*{Acknowledgments}
%\begin{acknowledgments}
\label{ackn}
We thank Paul Shapiro and Max Tegmark for helpful suggestions. This research was supported by the
DoE under Grant Nos. DE-FG02-95ER40896,
DE-FG02-04ER41308 and DE-FC02-94ER40818, by the NSF under grant Nos. PHY-0544278 and AST-0708176, by NASA under grant Nos.
NNX07AH09G and NNG04G177G, by Chandra grant No. SAO TM8-9009X, and by the Wisconsin
Alumni Research Foundation.

%\end{acknowledgments}

\end{document}